\documentclass{article}
\usepackage[preprint, nonatbib]{neurips_2022}

\usepackage[utf8]{inputenc} 
\usepackage[T1]{fontenc}    
\usepackage{hyperref}       
\usepackage{url}            
\usepackage{booktabs}       
\usepackage{amsfonts}       
\usepackage{nicefrac}       
\usepackage{microtype}      
\usepackage{xcolor}         
\usepackage{amsmath}
\usepackage{url}
\usepackage{amsfonts}
\usepackage{multirow}
\usepackage{booktabs}
\usepackage{bm}
\usepackage{array}
\newcolumntype{C}[1]{>{\centering\let\newline\\\arraybackslash\hspace{0pt}}m{#1}}

\usepackage{tikz}
\usetikzlibrary{positioning, decorations.pathmorphing}
\definecolor{nblue}{HTML}{88c0d0}
\definecolor{nwhite}{HTML}{eceff4}
\definecolor{nyellow}{HTML}{ebcb8b}
\definecolor{nred}{HTML}{bf616a}
\definecolor{ngreen}{HTML}{a3be8c}
\definecolor{npurple}{HTML}{b48ead}

\newcommand\T{\rule{0pt}{2.9ex}}       
\newcommand\B{\rule[-1.2ex]{0pt}{0pt}} 

\title{A Knowledge Graph-Enhanced Tensor Factorisation Model for Discovering Drug Targets}

\author{
  Cheng Ye,\enskip Rowan Swiers,\enskip  Stephen Bonner\enskip and Ian P Barrett\\
\rule[15pt]{0pt}{0pt}
{\normalsize Data Sciences and Quantitative Biology, Discovery Sciences, R\&D, AstraZeneca, Cambridge, UK}\\
\rule[15pt]{0pt}{0pt}
}

\begin{document}

\maketitle

\begin{abstract}
  The drug discovery and development process is a long and expensive one, costing over 1 billion USD on average per drug and taking 10-15 years. To reduce the high levels of attrition throughout the process, there has been a growing interest in applying machine learning  methodologies to various stages of  drug discovery and development in the recent decade, especially at the earliest stage – identification of druggable disease genes. In this paper, we have developed a new tensor factorisation model to predict potential drug targets (genes or proteins) for treating diseases. We created a three-dimensional data tensor  consisting of 1,048 gene targets, 860 diseases and 230,011 evidence attributes and clinical outcomes connecting them, using data extracted from the Open Targets and PharmaProjects databases. We enriched the data with gene target representations learned from a drug discovery-oriented knowledge graph and applied our proposed method to predict the clinical outcomes for unseen gene target and disease pairs. We designed three evaluation strategies to measure the prediction performance and benchmarked several commonly used machine learning classifiers together with Bayesian matrix and tensor factorisation methods. The result shows that incorporating knowledge graph embeddings significantly improves the prediction accuracy and that training tensor factorisation alongside a dense neural network outperforms all other baselines. In summary, our framework combines two actively studied machine learning approaches to disease target identification, namely tensor factorisation and knowledge graph representation learning, which could be a promising avenue for further exploration in data-driven drug discovery.
\end{abstract}

\section{Introduction}\label{sec:introduction}

The discovery and development of drugs is a complex process that aims to identify pharmalogical agents that are therapeutically effective for curing or treating diseases. According to the US Food and Drug Administration (FDA), on average, it takes approximately 12 years to get a newly discovered drug into the market and the cost is 1.3 billion USD \cite{fda}. There are many factors that contribute to the high cost of developing a drug: significant time and resources invested in target identification and the validation process, multiple early tests for a considerable number of promising compounds and complex in vitro and in vivo studies to understand the drug toxicity, to just name a few. Among all these factors, clinical research is the most critical and costly process. According to AstraZeneca, between the year 2005 and 2010, the success rate of preclinical projects in the company was 66\% while for Phase I, II, and III clinical projects, the rates dropped to 59\%, 15\% and 60\%, respectively \cite{cook2014lessons}. To raise the success rates of clinical projects and consequently, reduce the time and cost needed for developing a drug, it is imperative to develop an effective method to predict whether a candidate drug target is clinically promising for treating a disease on a large scale.

Traditionally, scientists identify new drug targets by discovering novel insights into a disease mechanism from research publications and lab experiments, which in turn allows scientists to design a drug molecule to activate or inhibit the gene target in order to cease or reverse the disease effects \cite{hughes2011principles}. Also, scientists investigate to reposition existing drugs to find unanticipated but beneficial effects against other diseases \cite{pushpakom2019drug}. Together, these processes have generated a large number of therapeutic hypotheses which are further supported by an enormous amount of both structured and unstructured evidence data. It is then of great interest to explore how we can better exploit such a wealth of information to identify and prioritise drug targets for diseases in an accurate and computationally efficient manner, using state-of-the-art data science technologies.

Machine learning is a natural candidate for this challenge. The gene target-disease pairs can be viewed as data points and their clinical trial outcomes as binary labels - success or failure. The biological evidence, from research publications, lab experiments, etc., that associates gene targets with diseases are used as features. However, there are several major difficulties in framing this as a standard binary classification problem. First, there are very few true positives – in Open Targets and PharmaProjects databases, less than 0.01\% gene target-disease pairs have approved drugs \cite{PharmaProjects, ochoa2021open}. Second, the feature space (biological evidence attributes) for pairs of gene targets and diseases is extremely sparse. In the September, 2020 version of Open Targets, there are in total 6,551,303 associations, which are extracted from 20 data sources, connecting 27,610 gene targets and 13,944 diseases. In other words, as high as 98.3\% of the gene target-disease pairs have no biological evidence at all. As a consequence, the traditional feature-in, prediction-out machine learning models might be incompetent in this context.

In this paper, motivated by the recent work \cite{yao2019predicting, wu2018neural}, we explore whether tensor factorisation can be used as a powerful tool for addressing the aforementioned challenges. The goal of our work is to generate novel therapeutic hypotheses for diseases of interest, i.e., finding potential gene targets that have a higher chance of being successfully used in  clinical trials and developed into effective medicines. Mathematically, a tensor is a higher order generalisation of a vector (first order tensor) and a matrix (second order tensor) in which each dimension corresponds to an axis or a mode. A three-mode tensor is therefore a natural representation to host the biological data in our analysis where the modes correspond to gene targets, diseases and their association evidence, respectively. The clinical outcomes are represented as an additional slice along the evidence mode. To predict the missing values in the clinical outcome slice, we have developed a novel tensor factorisation method that utilises gene target representations learned from a drug discovery-oriented knowledge graph as side information. The incorporated side information, specifically entity features, are shown to play a crucial role for predicting sparsely observed relations between gene targets and drugs \cite{simm2015macau}.   To the best of our knowledge, this is the first tensor factorisation framework that incorporates biomedical knowledge graphs for the purpose of predicting potential drug targets for diseases. To aid in reproducibility, we release all the processed public data sets and model training scripts\footnote{Code is available at: \url{https://github.com/AstraZeneca/kg-enhanced-tf-for-target-discovery}}.

The remainder of the paper is organised as follows. In Section \ref{sec:lit-review}, we review recent tensor factorisation applications in drug discovery and development. In Section \ref{sec:method}, we first describe our data collection and processing procedures and introduce how we use knowledge graphs and gene target embeddings in the developed framework. We then explain in detail our proposed tensor factorisation model.  We describe the benchmarking strategies and experimental setup in Section \ref{sec:experimental-setup}. In Section \ref{sec:results} we present the results of our experimental evaluation. Finally, we present our conclusions, discuss the contribution and limitations of our work, and highlight several future research directions.
\section{Literature Review}\label{sec:lit-review}

In recent decades, tensor factorisation has found many useful applications in a wide range of machine learning areas, such as recommender systems, image processing and computer vision, and natural language processing \cite{anisimov2014semantic, haeffele2014structured, van2009non, park2017rectime, zheng2016topic}. Such success has led to a growth in the use of tensor factorisation for addressing biomedical challenges in recent years. For instance, Chen and Li \cite{chen2018drugcom} proposed a tensor decomposition model to predict the therapeutic benefits of drug combinations. In particular, a three-mode tensor was constructed to represent comprehensive relationships between drugs and diseases as well as their similarity information. The molecular mechanisms of drug synergy can thus be well revealed via simultaneously factorising the coupled tensor and similarity matrices. The authors later applied the technique to model relational drug-gene target-disease interactions and successfully showed that the model outperforms some competitive methods at predicting drug mechanism of action \cite{chen2019modeling}. As we previously discussed, data sparsity and scalability remains an important issue in this field, and recently,  Macau, a powerful, flexible and scalable Bayesian multi-relational factorisation method for heterogeneous data has been developed  to address this challenge \cite{simm2015macau}. One unique advantage of Macau is that it enables the incorporation of side information, specifically entity and relation features, which is essential for complex tasks such as predicting sparsely observed drug-protein activities. More recently, Yao et al. \cite{yao2019predicting} have applied Macau to the problem of predicting clinical outcomes of therapeutic hypotheses and have also tested the performance of several other machine learning models. The result suggested that tensor factorisation is comparable or better than the baseline methods in various cross-validation scenarios.

The importance of gene target side information in matrix and tensor factorisation has been discussed by Piro and Di Cunto \cite{piro2012computational}. In the paper, they have examined a number of disease gene prioritisation tools that utilise different types of evidence including text-mining of biomedical literature, functional annotations, pathways and ontologies, phenotype relationships, intrinsic gene properties, sequence data, protein–protein interactions, regulatory information, orthologous relationships and gene expression information. Generally speaking, most existing methods incorporate gene target side information by either directly merging data from different sources or applying techniques such as principal component analysis (PCA) to evidence data \cite{natarajan2014inductive, vitsios2020mantis}. However, such ways of using side information do not fully exploit the interconnections between genes and other biological entities and dimensionality reduction technologies such as PCA cannot capture the essential gene representations in the high-dimensional, heterogeneous data. To address this challenge, we utilise a biomedical knowledge graph, Hetionet \cite{himmelstein2017systematic}, which is an integrative network of biomedical knowledge centered around genes and diseases assembled from different databases, and use gene representations automatically learned from the graph (known as knowledge graph embeddings \cite{wang2017knowledge, wang2014knowledge}) as the gene target side information in our proposed model. We show that by connecting different biological resources via a well-structured knowledge graph and by learning gene representations which contain essential information not only from gene-gene interactions and gene-disease associations but the broader biomedical knowledge field, the prediction performance can be improved.
\section{Method}\label{sec:method}

\subsection{Task Overview}

The task in this paper is, given a pair of gene target and disease, to predict whether the gene, which is targeted by a developed drug for treating the disease, can lead to success in a clinical trial. Mathematically, we aim to develop a function $F$ that takes input of any disease $D_i$ and gene target $T_j$ in the dataset, and produces a score $F(D_i, T_j) \in [0, 1]$, indicating the probability of this gene target-disease hypothesis being developed into successful medicines. The details of obtaining the data, and the development of the methodology are described in the following subsections.

\subsection{Data Collection and Processing}

We have created a dataset which combines clinical outcomes from the commercial database CiteLine PharmaProjects~\cite{PharmaProjects} and evidence from the open-source database Open Targets~\cite{ochoa2021open} between gene targets and diseases (this is detailed in Table~\ref{tab:opentargets}). PharmaProjects has curated lists of clinical trials specifying the drugs being tested, the targeted proteins and the diseases being tested against. We joined these two datasets together to create a mapping of diseases to gene targets. We defined that the clinical outcome of a gene target and a disease is positive if there has been at least one successfully completed clinical trial resulting in a launched drug for the pair in the PharmaProjects database. On the other hand, the clinical outcome is negative if there is no clinical success and at least one failed clinical trial. Open Targets has seven different evidence types, each with an association score, for each gene target-disease pair. We therefore used Open Targets to create the evidence data between targets and diseases.

\begin{table*}[ht!] %
    \centering
    \caption{Gene Target-disease evidence types in Open Targets.}\label{tab:opentargets}
    \renewcommand{\arraystretch}{1.2}
    \begin{tabular}{p{0.2\textwidth} C{0.35\textwidth} C{0.35\textwidth}}
        \toprule
        \textbf{Evidence Type}       & \textbf{Association Score Sources}                                                                                                  & \textbf{Description}\T\B                                                                                                                                                              \\
        \midrule \midrule

        Genetic Associations         & ClinVar (EVA), PheWAS Catalog, Gene2Phenotype, Genomics England PanelApp, Open Targets Genetics Portal, UniProt literature, ClinGen & Curated and calculated associations between genetic variants and mendelian-inherited diseases, common diseases and phenotypes/traits.                                                 \\

        \midrule

        Somatic Mutations            & Cancer Gene Census, ClinVar somatic (EVA), IntOGen                                                                                  & Somatic mutations noted in cancers and other diseases.                                                                                                                                \\

        \midrule

        Drugs                        & ChEMBL                                                                                                                              & Database of known drugs and other pharmacological agents linked to bioactivity assay data and disease (where applicable).                                                             \\

        \midrule

        Pathways and Systems Biology & Reactome, Sysbio, SLAPenrich, PROGENy, Project Score (CRISPR)                                                                       & Gene sets comprised from curated biological pathways, and oncology-centric resources (CRISPR gene function screens and pathway activities inferred from gene expression data).        \\

        \midrule

        RNA Expression               & Expression Atlas                                                                                                                    & Gene expression data from different biological conditions (e.g.\ normal and disease).                                                                                                 \\

        \midrule

        Text Mining                  & EuropePMC                                                                                                                           & Co-occurrences between gene target and disease terms, text mined from PubMed and PubMedCentral.                                                                                       \\

        \midrule

        Animal models                & PhenoDigm                                                                                                                           & Associations between gene targets and diseases, inferred from similarities between human disease characteristics and phenotypes observed in mouse models where the gene is perturbed. \\

        \bottomrule
    \end{tabular} 
\end{table*}

One major challenge occurring when combining the two data sets was that PharmaProjects uses the MeSH Ontology~\cite{lipscomb2000medical} for diseases while Open Targets uses the Mondo Ontology \cite{mungall2017monarch}. In general, disease ontologies are widely used for annotation, integration and analysis of disease-related biological data, and their range and diversity are often high due to different areas they are used in \cite{kurbatova2021disease}. In our case, this means that we have multiple disease terms (IDs) as well as hierarchies originating from the two disease ontologies, MeSH and Mondo, for describing a same disease. To address this challenge, we used EMBL-EBI Ontology matching tool OXO \cite{oxo} to obtain cross-reference information (ontological matching results) in disease ontologies and connect the MeSH and Mondo terms.

\begin{figure}[!ht]
    \centering
    \includegraphics[width=0.35\textwidth]{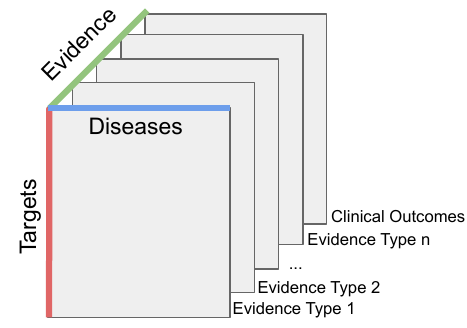}
    \caption{A tensor representation of the biological dataset}\label{fig:tf-bio}
\end{figure}

We finally obtained a three-dimensional tensor in which the first two dimensions correspond to 1,048 gene targets and 860 diseases, respectively and the third dimension presents their evidence and clinical outcomes (230,011 in total), which is depicted in Figure~\ref{fig:tf-bio}. The tensor is binary, i.e., all of its entries are either 0 or 1, indicating whether or not a gene target-disease pair is linked by some evidence type or by the outcome of clinical trials. The density of the tensor is 3.19\%.

\subsection{Knowledge Graph and Gene Target Embeddings}

Generally speaking, a knowledge graph is an integrative network representation of structured and unstructured knowledge, which provides an easier way of capturing complex semantic relationships than conventional databases. For instance, in a biomedical knowledge graph, various biological entities are modelled as nodes, whilst millions of complex biomedical relationships are modelled as edges (or links) connecting nodes \cite{bonner2021review}. In recent years, many researchers and pharmaceutical companies have been building biomedical knowledge graphs to address various drug discovery challenges, e.g., Hetionet \cite{himmelstein2017systematic}, PharmKG \cite{zheng2020pharmkg} and Rosalind's knowledge graph \cite{paliwal2020preclinical}.

In this paper, we make use of Hetionet, which is one of the earliest integrative biological networks centred around genes and diseases and now forms the core of many more recently developed disease knowledge graphs, and learn effective feature representations (also known as embeddings) of gene targets using various knowledge graph embedding techniques. These techniques are: Node2vec, a homogeneous graph embedding method that groups nodes with similar neighbourhoods together \cite{grover2016node2vec}; Metapath2vec, a heterogeneous graph embedding method that groups nodes of the same type with similar neighbourhoods together \cite{dong2017metapath2vec}; Graph Variational Auto-Encoder (GVAE), a homogeneous graph-based embedding approach that tries to predict links between nodes by applying a convolutional operation to latent variables of node neighbours \cite{kipf2016variational}; Relational Graph Convolutional Networks (RGCN), an extended class of GCNs, which is developed specifically for knowledge base completion tasks (link prediction and entity classification) on  multi-relational data of realistic knowledge bases \cite{Schlichtkrull2017modelling}. All the algorithms were set to use embedding dimension of 32 and were trained using the default parameters reported in the respective paper.

\begin{figure*}[!t]
    \centering
    \resizebox{0.85\textwidth}{!}{
        {\tiny
                \begin{tikzpicture}

                    \node[circle, draw, very thick, minimum size=1cm, fill=nyellow!50, align=center] at (-2, -0.5) (bp1) {Biological \\ Process};
                    \node[circle, draw, very thick, minimum size=1cm, fill=nyellow!50, align=center] at (1, -4.5) (bp2) {Biological \\ Process};
                    \node[circle, draw, very thick, minimum size=1cm, fill=ngreen!50] at (-1, 2) (p) {Pathway};
                    \node[circle, draw, very thick, minimum size=1cm, fill=nblue!50] at (0, -2) (g1) {Gene};
                    \node[circle, draw, very thick, minimum size=1cm, fill=nblue!50] at (3, -3) (g2) {Gene};
                    \node[circle, draw, very thick, minimum size=1cm, fill=nblue!50] at (1.5, 0.7) (g3) {Gene};
                    \node[circle, draw, very thick, minimum size=1cm, fill=nblue!50] at (4.2, -0.5) (g4) {Gene};
                    \node[circle, draw, very thick, minimum size=1cm, fill=npurple!50] at (6, -2.5) (d1) {Disease};
                    \node[circle, draw, very thick, minimum size=1cm, fill=npurple!50] at (5.5, 1.5) (d2) {Disease};
                    \node[circle, draw, very thick, minimum size=1cm, fill=npurple!50] at (3.5, 2.5) (d3) {Disease};
                    \node[circle, draw, very thick, minimum size=1.1cm, fill=nred!50] at (4.8, -5) (c) {Drug};

                    \draw[-, line width=2pt, color=nblue!90] (g1) to  node[below, sloped, text width=3cm, text centered,  color=black]{\tiny Regulates}(g2);
                    \draw[-, line width=2pt, color=nblue!90] (g1) to node[below, sloped, text width=3cm, text centered,  color=black]{\tiny Interacts}(g3);
                    \draw[-, line width=2pt, color=nblue!90] (g2) to node[below, sloped, text width=3cm, text centered,  color=black]{\tiny Interacts}(g3);
                    \draw[-, line width=2pt, color=nblue!90] (g3) to node[below, sloped, text width=3cm, text centered,  color=black]{\tiny Covaries}(g4);

                    \draw[-, line width=2pt, color=nyellow!90] (bp1) to node[below, sloped, text width=3cm, text centered, color=black]{\tiny Participates}(g1);
                    \draw[-, line width=2pt, color=nyellow!90] (g1) to node[below, sloped, text width=3cm, text centered, color=black]{\tiny Participates}(bp2);
                    \draw[-, line width=2pt, color=nyellow!90] (g2) to node[below, sloped, text width=3cm, text centered, color=black]{\tiny Participates}(bp2);

                    \draw[-, line width=2pt, color=ngreen!90] (p) to node[below, sloped, text width=3cm, text centered, color=black]{\tiny Participates}(g3);

                    \draw[-, line width=2pt, color=npurple!90] (d3) to node[below, sloped, text width=3cm, text centered, color=black]{\tiny Associates}(g3);
                    \draw[-, line width=2pt, color=npurple!90] (d2) to node[below, sloped, text width=3cm, text centered, color=black]{\tiny Upregulates}(g4);
                    \draw[-, line width=2pt, color=npurple!90] (d1) to node[below, sloped, text width=3cm, text centered, color=black]{\tiny Downregulates}(g4);
                    \draw[-, line width=2pt, color=npurple!90] (d1) to node[below, sloped, text width=3cm, text centered, color=black]{\tiny Downregulates}(g2);

                    \draw[-, line width=2pt, color=nred!90] (c) to node[below, sloped, text width=3cm, text centered, color=black]{\tiny Binds}(g2);
                    \draw[-, line width=2pt, color=nred!90] (c) to node[below, sloped, text width=3cm, text centered, color=black]{\tiny Treats}(d1);

                    \coordinate(a1) at (7, -0.5);
                    \coordinate(a2) at (10, -0.5);
                    \draw[-stealth, line width=3pt, color=black, label=below:text]  (a1) to node[below, sloped, text width=3cm, text centered, yshift=-4pt]{\small Knowledge Graph \\ Embeddings}(a2);

                    \draw[draw=black] (12, -4) rectangle ++(8,7);

                    \node[circle, draw, very thick, minimum size=0.5cm, fill=nblue!50] at (16, 0.7) (g) {Gene};
                    \node[circle, draw, very thick, minimum size=0.5cm, fill=nblue!50] at (17, -0.5) (g) {Gene};
                    \node[circle, draw, very thick, minimum size=0.5cm, fill=nblue!50] at (17.5, 0.5) (g) {Gene};
                    \node[circle, draw, very thick, minimum size=0.5cm, fill=nblue!50] at (16, -0.3) (g) {Gene};

                    \node[circle, draw, very thick, minimum size=0.5cm, fill=nwhite!90] at (13, 2) (d) {Disease};
                    \node[circle, draw, very thick, minimum size=0.5cm, fill=nwhite!90] at (12.5, 1) (d) {Disease};
                    \node[circle, draw, very thick, minimum size=0.5cm, fill=nwhite!90] at (13.9, 0.4) (d) {Disease};

                    \node[circle, draw, very thick, minimum size=0.5cm, fill=nwhite!90] at (13.5, -3) (d) {Pathway};

                    \node[circle, draw, very thick, minimum size=0.5cm, fill=nwhite!90, align=center] at (18, -2.5) (d) {Bio \\ Process};
                    \node[circle, draw, very thick, minimum size=0.5cm, fill=nwhite!90, align=center] at (19, -1) (d) {Bio \\ Process};

                    \node[circle, draw, very thick, minimum size=0.5cm, fill=nwhite!90] at (19, 2) (d) {Drug};

                    \node[] at (16,-4.5) {Embedding Dimension 1};
                    \node[rotate=-270] at (20.5,-0.5) {Embedding Dimension 2};

                \end{tikzpicture}
            }}
    \caption{A knowledge graph schema and associated embeddings.}\label{fig:tf-schema}
\end{figure*}
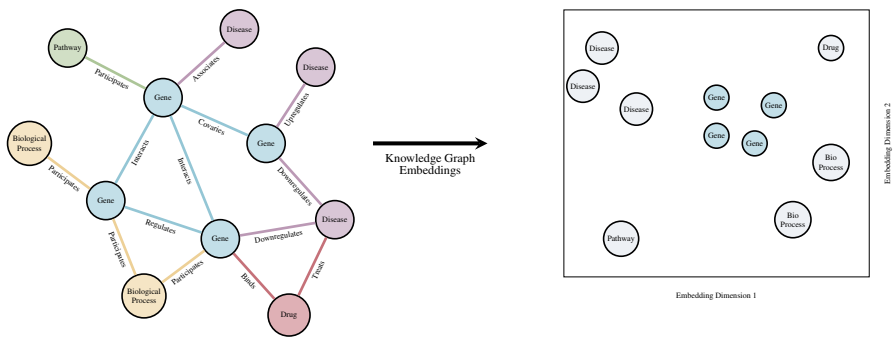

Hetionet contains 47,031 biological entities of 11 types (Gene, Disease, Compound, Pathway, Anatomy etc.) and 2,250,197 relationships of 24 types (Disease-associates-Gene, Compound-treats-Disease, Gene-participates-Pathway, etc.) integrated from 29 public databases. The graph models the complex interactions between genes and other genes, diseases, compounds, pathways, biological processes and many other entities, so the learned embeddings are able to capture  interactive characteristics of genes in the biomedical field \cite{bonner2021understanding} (see Figure~\ref{fig:tf-schema}). We therefore used the gene embeddings to enrich the gene target information in our data set and treated them as side information in the tensor factorisation framework to help improve the prediction accuracy.

\subsection{Tensor Factorisation with Knowledge Graph Embeddings}

\begin{figure*}[!t]
    \centering
    \includegraphics[width=0.8\textwidth]{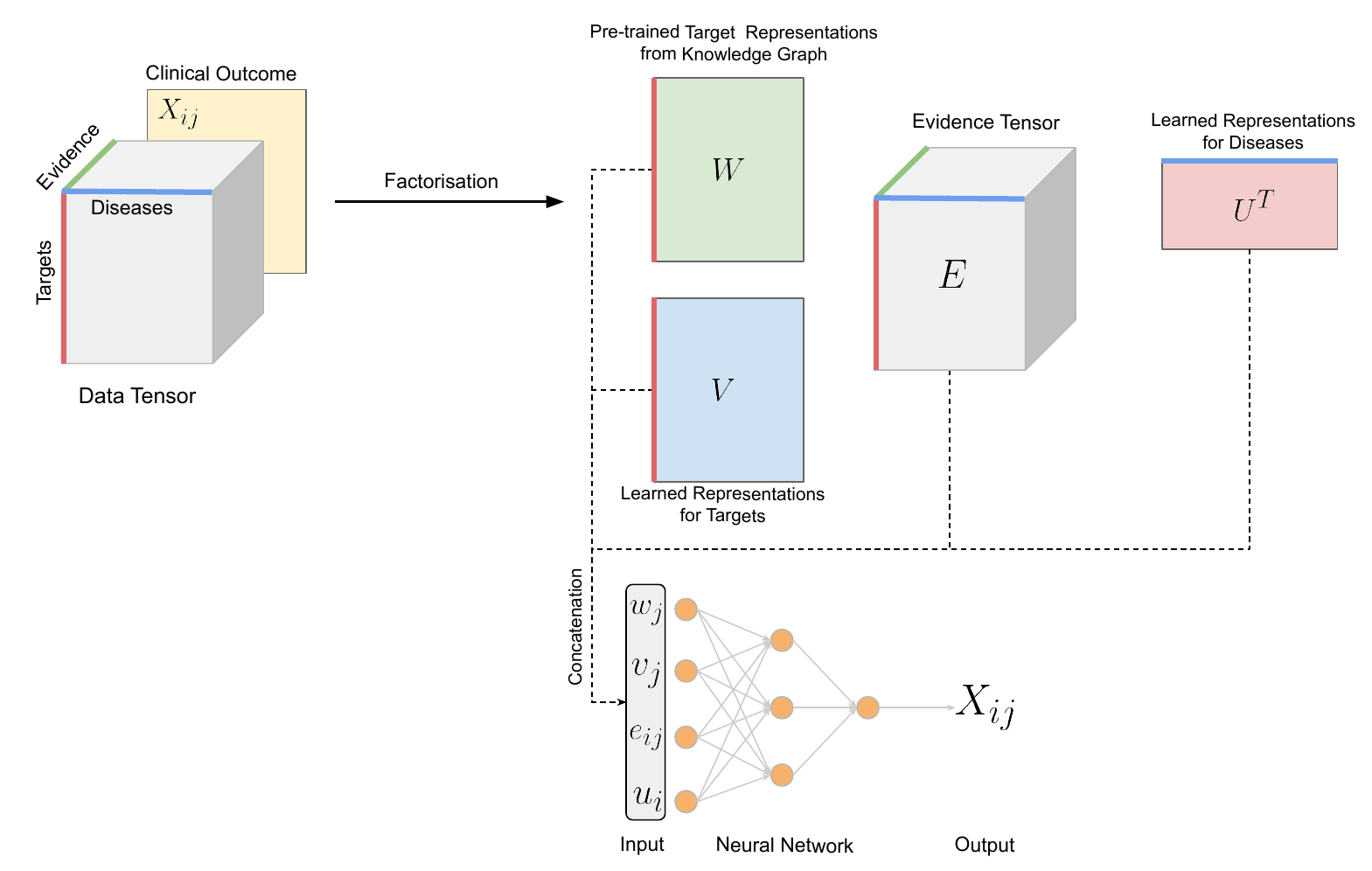}
    \caption{Schema of the proposed model.}\label{fig:model}
\end{figure*}

To predict the missing values in the clinical outcome matrix in the data tensor, we adopted a tensor factorisation model which factorises the data tensor into a core evidence tensor and two latent gene target and disease matrices that need to be learned, as depicted in Figure \ref{fig:model}. These together with the gene target representation matrix learned from the knowledge graph were then fed into a dense neural network to generate the output, which will be the predicted value for the clinical outcome of a gene target-disease pair.

We denote $D = \{D_1,...,D_n\}$ the set of diseases and $T=\{T_1,\ldots,T_n\}$ the set of gene targets present in the data set and the association score of evidence type $k$ between disease $D_i$ and gene target $T_j$ is represented as $E_{ijk}\in\{0,1\}$. Similarly, the clinical outcome between disease $D_i$ and gene target $T_j$ is denoted as $X_{ij}\in\{0,1\}$. Let the learned latent variables for diseases and gene targets, and the gene target knowledge graph embeddings be $U$, $V$, $W$, respectively. Then the probability of a successful clinical outcome between disease $D_i$ and gene target $T_j$ is calculated as follows:

\begin{equation}
    P(X_{ij}=1|\mathbf{e_{ij}}, \mathbf{u_i}, \mathbf{v_j}, \mathbf{w_j}) = f(concat( \mathbf{e_{ij}}, \mathbf{u_i}, \mathbf{v_j}, \mathbf{w_j} )),
\end{equation}
where $\mathbf{e_{ij}}, \mathbf{u_i}, \mathbf{v_j}, \mathbf{w_j}$ are the evidence vector between $D_i$ and $T_j$, latent variables for $D_i$, latent variables for $T_j$, and gene target representations learned from the knowledge graph for $T_j$, respectively. These vectors are concatenated together before being input into the function. We define the function $f$ to be a feed-forward neural network with 2 dense hidden layers of 256 and 64 neurons, both using the ReLU (Rectified Linear Unit)  activation function. The output layer uses a linear activation function. It is important to notice that the proposed training paradigm is a two-stage framework, i.e., tensor factorisation first, followed by the neural network classification.

The dimensionality of the gene target and disease parameter matrices is $V \in \mathbb{R}^{|T|,d}$ and $U \in \mathbb{R}^{|D|,d}$, where $d$ is the size of the embedding for each entity – set to 32 for this work. We have also explored other dimensionality numbers but found no significant difference in the result. Putting all these together, these matrices contribute 61,056 learnable parameters to the model. The neural network, when using a 32-dimensional gene target embeddings, has 43,240 parameters. This results in the final total model size of just over 100,000 learnable parameters. These parameters were optimised using the Adagrad \cite{ward2019adagrad} variant of Stochastic Gradient Descent (SGD) with Mean Squared Error (MSE) as the loss function.  The model was trained for 100 epochs, using a batch size of 2,000 and a learning rate of 0.05.

We argue that the main advantage of this proposed technique is that the dense neural network allows for complex interactions between the Open Targets evidence, the latent variables and the pre-trained knowledge graph embeddings. In this way the embeddings from the knowledge graph can have non-linear interactions with the prediction outcome and can interact with both the disease and the gene target latent variables. This would not be happening if the embeddings were used as side information in algorithms such as Bayesian matrix and tensor decompostion models, e.g., Macau \cite{simm2015macau}.

\subsection{Modelling Choice}
Generally speaking, there are two approaches to tensor decomposition, namely CANDECOMP/PARAFAC decomposition (CPD), which factorises a tensor into a sum of rank-one tensors and the Tucker decomposition, which can be considered as a higher-order form of principal component analysis \cite{kolda2009tensor}. Notably, our proposed tensor factorisation model lies within the Tucker2 model family as it factorises the data tensor into a full core (the evidence tensor) $E$ and two component matrices for gene targets $V$ and diseases $U$, respectively, with the component matrix for the evidence mode missing. It is in fact identical to joint matrix factorisation and is widely used in association with graph information in many collaborative filtering applications including \cite{gu2010weighted, yao2014dual, chen2016fascinate}.

The main reason behind our choice of using a Tucker model instead of CPD is the presence of the core tensor in the Tucker model, whose element determines the weight of the outer product between the corresponding factors from each  component matrix \cite{kolda2009tensor}, which allows us to make full use of the gene target-disease evidence array from Open Targets. Furthermore, we specifically selected a Tucker2 model is because the Tucker3 model is a true three-way decomposition model in which it explicitly establishes a relationship between factors in the three modes spanned by the data array. However, in our task, we only have two meaningful modes for compression -- gene targets and diseases, and leave the third mode -- the evidence -- uncompressed. Thus, our model is essentially a Tucker2 model which uses two component matrices and a full core.

\section{Experimental Setup}\label{sec:experimental-setup}

\subsection{Datasets Overview}

The three-dimensional tensor used in the experiments, which was constructed from the Open Targets and PharmaProjects databases, contains 1,048 gene targets, 860 diseases and 230,011 evidence attributes and clinical outcomes connecting them. The drug discovery-oriented knowledge graph Hetionet, which is used to enrich the side information of  gene targets, consists of 47,031 biological entities and 2,250,197 relationships integrated from 29 public databases.

\subsection{Evaluation Strategies}

We used three cross-validation strategies to assess the accuracy of predicting the unseen clinical outcomes of gene target and disease pairs. The first randomly split all the gene target-disease pairs into five folds. We trained the models on four folds and tested on the remaining fold. This is the standard cross-validation procedure; however, it could result in the case that gene target-disease pairs that share the same gene target, or the same disease are assigned to different folds, and thus introduce bias when measuring the performance. This is because if one gene target has been clinically successful for a particular disease, then it is likely to be successful for another closely related disease. Similarly, if a disease is treated well by a particular gene target then it is likely to be treated well by another similar gene target.

To overcome this issue, we designed a second strategy where we first randomly assigned gene targets into one of five groups. Gene target-disease pairs are then put into the group the gene target is assigned to. In this way, the gene target-disease pairs that share the same gene target are guaranteed to be in the same group. We then again trained the models on four groups and tested on the remaining group. Lastly and similarly, we split all diseases into several classes depending on which part of the human body the disease affects. The gene target-disease pairs are then put into the class the disease is assigned to.

The similar evaluation strategies have been used to address other biomedical challenges. For instance, in \cite{Lim2016largescale}, to measure the performance of predicting chemical-protein associations, the authors have divided the dataset into several categories based on the connectivity of known chemical-protein associations and the degree of uniqueness of the chemicals and then performed cross-validation accordingly. Moreover, in \cite{Li2019learning}, given a three-dimensional binary tensor of diseases, genes and chemicals, the authors have assessed the performances of predicting the disease-gene-chemicals triples by performing fiber-wise evaluation and slice-wise evaluation for disease, gene and chemical tensor respectively.

\subsection{Baseline Models}

We applied the following classic machine learning classifiers to the data as baselines. Logistic regression, a simple linear model using the Scikit-Learn implementation \cite{pedregosa2011scikit}; k-nearest neighbours classifier using the Scikit-Learn implementation \cite{pedregosa2011scikit}; Random forest, an ensemble method using the Scikit-Learn implementation \cite{pedregosa2011scikit}; MLP, a multi-layer perceptron neural network method using the Scikit-Learn implementation \cite{pedregosa2011scikit}; Gradient boosting machine using the XGBoost implementation \cite{chen2016xgboost}.

We also compared our method with matrix factorisation (applying on the gene target-disease clinical outcome matrix only) and tensor factorisation (applying on the full tensor data) baseline models. Following \cite{yao2019predicting}, we selected a framework called Macau \cite{simm2015macau}, which is  a powerful and flexible method for factorisation of heterogeneous data, and was originally designed for drug-protein interaction prediction. In particular, Macau serves as a Bayesian probabilistic framework for both matrix and tensor decomposition, and its advantages over other similar methods include that it is able to factorise a wide range of data models, incorporate features for any entities and relations and scale up to millions of entity instances, hundred millions of observations.

\subsection{Performance Metrics}

Two metrics have been used to assess the prediction performance of the evaluated methods: Area Under Receiver Operator Curve (AUROC) and Area Under Precision-Recall Curve (AUPRC). The former is normally used to measure how well a classifier can distinguish between a positive sample and a randomly selected negative sample in a binary classification problem while the latter assesses the probability that if a positive sample is selected from the ranked list produced by a method, all samples ranked above it are positive.
\section{Results}\label{sec:results}

\begin{figure*}[!t]
    \centering
    \includegraphics[width=0.99\textwidth]{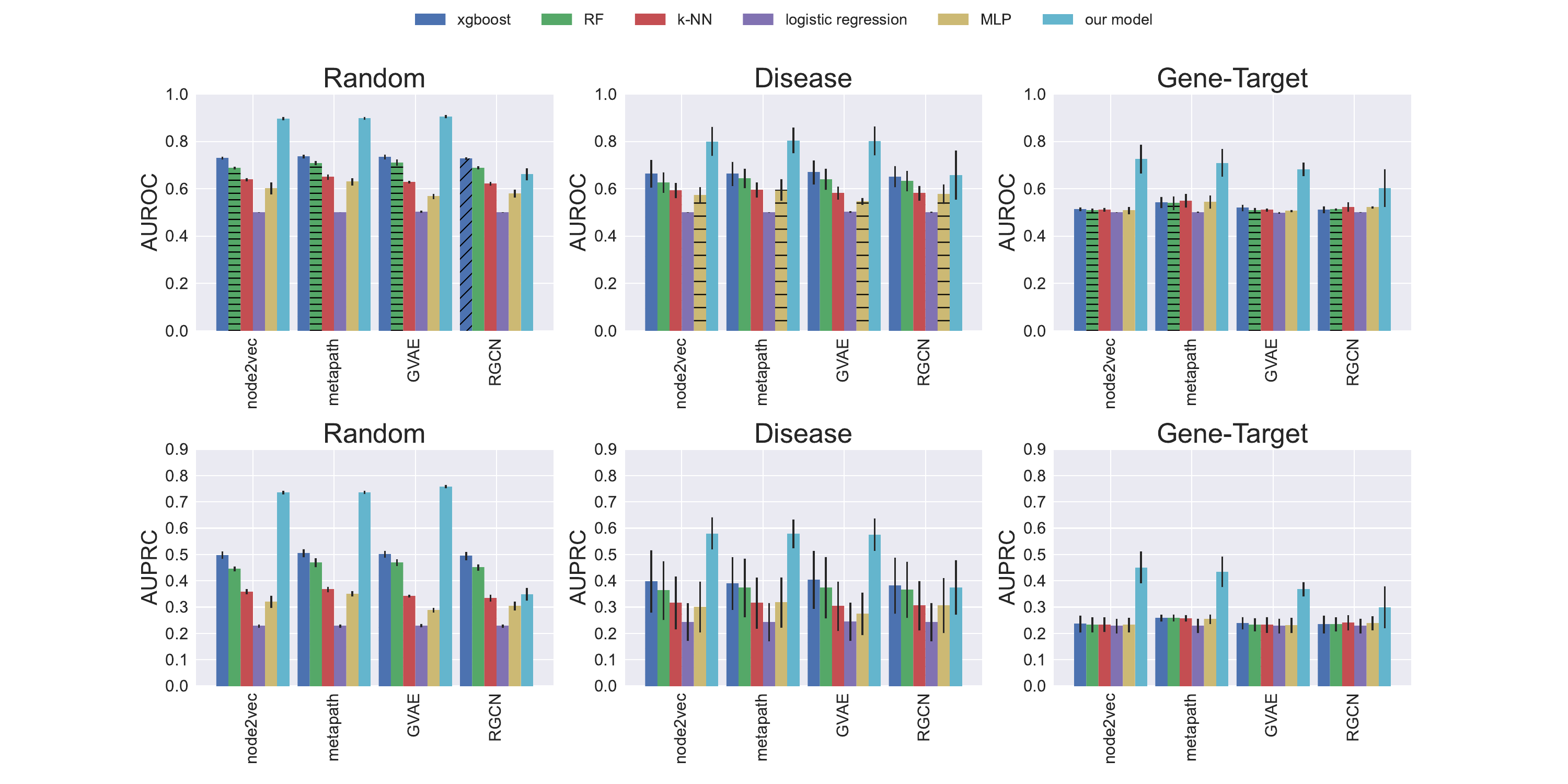}
    \caption{Results comparing our proposed method with various other machine learning classifiers when using representations from the various approaches. Results are presented across different splits of the dataset as described in Section 4.2.}
    \label{fig:result1}
\end{figure*}

\subsection{Evaluating Baseline Models}

In this experiment, we evaluated the prediction performance of our proposed tensor factorisation method by comparing with baseline models using both gene embeddings and the evidence tensor.

Figure~\ref{fig:result1} reports the mean (with standard deviation) over the two performance metrics using all three cross-validation strategies detailed in Section 4.2. The most striking trend in the figure is that in all three evaluation scenarios, our method outperforms all the baseline models by a significant margin (over 0.2 in AUROC and over 0.3 in AUPRC, on average), demonstrating the superiority of the tensor factorisation method. Several other important observations can be made. First, as we suspected, it is indeed much easier to make predictions using the random split benchmark than the disease or gene target split benchmark. This is because in the random split, the algorithms are more likely to have trained on similar pairs of gene targets and diseases, which tend to have a higher success rate if one of them is a success. We believe that the disease split and gene target split scenarios may give a more realistic indication of success chance and are thus more relevant. Second, in the disease and random split cases, there is a significant improvement in performance by including some type of gene target-level embeddings. This could be because the embeddings uniquely identify the gene targets and therefore allow the classifiers to “memorise” the successful drug targets for a particular disease and make accurate prediction for other similar gene targets for that disease.

Comparing the performance of the various graph embedding approaches more closely, some interesting observations can be made. For example, on the disease and random splits, it can be seen that the choice of the graph embedding has little impact on the overall predictive performance - apart from the RGCN embeddings, which results in significantly worse performance overall.  On the gene target split, both the Metapath2vec and Node2vec embeddings outperform the others.

\subsection{Evaluating Matrix and Tensor Factorisation Methods}

In this experiment, we compared our proposed tensor factorisation method with Macau Bayesian probabilistic matrix factorisation and  tensor factorisation~\cite{simm2015macau}. The gene target representations in our method are the Metapath2vec embeddings. We again used all three evaluation strategies, and the results are highlighted in Table \ref{tab:model_macau}. It demonstrates that in the disease and gene target split scenarios, our model clearly outperforms both the Bayesian matrix and tensor factorisation methods, which shows how incorporating knowledge graph embeddings into the tensor factorisation can lead to better overall predicative performance. Our approach is also slightly better than the Macau matrix and tensor factorisation on the random data split.

\begin{table*}[h!]
    \centering
    \caption{Performance of our method (using the Metapath2Vec representations) compared with Macau matrix (MF) and tensor factorisation (TF) across three dataset splits.}\label{tab:model_macau}
    \begin{tabular}{l l c c c c c }
        \toprule
        \textbf{Metric}        & \textbf{Dataset Split} & \multicolumn{3}{c}{\textbf{Approaches}}\T\B                                                             \\
        \midrule \midrule
                               &                        & Macau (TF)                                  & Macau (MF)                 & Our Model (MP2V)\B           \\
        \cline{3-5}

        \multirow{3}{*}{AUROC} & Random                 & 0.912\(\pm\)0.004                           & \textbf{0.932\(\pm\)0.004} & 0.893\(\pm\)0.007\T          \\
                               & Disease                & 0.658\(\pm\)0.052                           & 0.664\(\pm\)0.054          & \textbf{0.801\(\pm\)0.065}   \\
                               & Gene-Target            & 0.501\(\pm\)0.001                           & 0.501\(\pm\)0.001          & \textbf{0.721\(\pm\)0.044}\B \\
        \midrule
        \multirow{3}{*}{AUPRC} & Random                 & 0.810\(\pm\)0.006                           & \textbf{0.847\(\pm\)0.004} & 0.727\(\pm\)0.018\T          \\
                               & Disease                & 0.389\(\pm\)0.101                           & 0.403\(\pm\)0.098          & \textbf{0.575\(\pm\)0.128}   \\
                               & Gene-Target            & 0.228\(\pm\)0.027                           & 0.227\(\pm\)0.027          & \textbf{0.428\(\pm\)0.031}\B \\

        \bottomrule
    \end{tabular}
\end{table*}
\subsection{Abalation}

To gain deeper insights into the roles of different components in the proposed tensor factorisation framework, we conducted an ablation study to assess the importance of each of the three sources of information used in our model: the learned representations for gene targets and diseases (matrices \(V\) and \(U\) in Figure~\ref{fig:model}), the evidence tensor ($E$) and the pre-trained gene target representations taken from Hetionet (matrix $W$). Table~\ref{tab:abalation} highlights the results. The most important conclusion is that the full model (with all features) presents the best performance, suggesting that our method is able to effectively utilise various sources of information to produce better predictions. On a closer look, the model relies particularly on both the learned representations and the embeddings provided for the gene targets from Hetionet. Indeed, these latent representations and knowledge graph embeddings enable the model to have comparable performance even when no evidence tensor is provided -- highlighting that the model can make good use of the side information resided in the graph.

\begin{table}[h!]
    \centering
    \small
    \caption{Ablation study over information sources.}\label{tab:abalation}
    \begin{tabular}{l l c c c c c }
        \toprule
        \textbf{Metric}        & \textbf{Dataset Split} & \multicolumn{3}{c}{\textbf{Omitted}}\T\B                                                                \\
        \midrule \midrule
                               &                        & Learned Representations                  & Evidence Tensor   & Target Embeddings & None\B               \\
        \cline{3-6}

        \multirow{3}{*}{AUROC} & Random                 & 0.584\(\pm\)0.056                        & 0.928\(\pm\)0.004 & 0.834\(\pm\)0.021 & 0.893\(\pm\)0.007 \T \\
                               & Disease                & 0.565\(\pm\)0.099                        & 0.716\(\pm\)0.044 & 0.672\(\pm\)0.077 & 0.801\(\pm\)0.065    \\
                               & Gene-Target            & 0.595\(\pm\)0.060                        & 0.641\(\pm\)0.019 & 0.551\(\pm\)0.065 & 0.721\(\pm\)0.044 \B \\
        \midrule
        \multirow{3}{*}{AUPRC} & Random                 & 0.300\(\pm\)0.044                        & 0.800\(\pm\)0.013 & 0.584\(\pm\)0.044 & 0.727\(\pm\)0.018    \\
                               & Disease                & 0.305\(\pm\)0.112                        & 0.449\(\pm\)0.078 & 0.402\(\pm\)0.098 & 0.575\(\pm\)0.128    \\
                               & Gene-Target            & 0.308\(\pm\)0.059                        & 0.347\(\pm\)0.020 & 0.273\(\pm\)0.056 & 0.428\(\pm\)0.031\B  \\

        \bottomrule
    \end{tabular} 
\end{table}

\subsection{Training Times}

All models were trained on an Intel(R) Xeon(R) Gold 6252 CPU @ 2.10GHz with 32 cores. The training time (see Table \ref{tab:table3}) of our model is approximately 50\% faster than Macau tensor factorisation, but it is slower than both the baselines and Macau matrix factorisation. However we argue that the predictive performance gain of our model outweighs the computational burden.

\begin{table}[]
    \centering
    \caption{Model training times}\label{tab:table3}
    \begin{tabular}{c c}
        \toprule
        \textbf{Method}            & \textbf{Training time (min:sec)} \\
        \midrule \midrule
        Our model                  & 11:27                            \\
        Macau matrix factorisation & 01:02                            \\
        Macau tensor factorisation & 22:57                            \\
        Baseline models            & 07:35                            \\
        \bottomrule
    \end{tabular}
\end{table}

\subsection{Evaluating top gene target-disease predictions}

\textbf{Low Data Diseases.} In order to gain some subjective insight into our model predictions, we examined three diseases that have only one gene target in the source data and examined the top 3 novel predicted gene targets for each of them (see Table \ref{tab:disease_low}). These are genes for which no evidence was found linking them to the disease in the original dataset. Interestingly, some of the predicted relationships would appear to have some initial plausibility. For example, gene TYMP is shown to be linked to breast adenosis and breast fibrocystic disease in a number of publications \cite{ozyilkan200015, eskelinen1992prospective, willsher1995serum, sutterlin1997clinical}. Moreover, the expression of gene FLT1, also known as VEGFR-1, a member of the vascular endothelial growth factor receptor (VEGFR) family, was shown to be significantly increased in breast cancer tumour tissue compared with healthy breast tissue in the patients with benign breast disease in multiple articles \cite{liu2003expression, caine2007comparison, srabovic2013vascular}. On a closer examination, although there is no direct textual evidence supporting the top predicted target SLC12A3, very recent research \cite{li2021solute} suggests that another member of solute carrier family 12, SLC12A8,  was indeed up-regulated in breast carcinoma and its knock-down and over-expression suppressed and elevated the viability, invasiveness and motility of breast carcinoma cells, respectively. Additional pathway analysis reveals that SLC12A8 plays a promoting effect in breast carcinoma by activating the TLR/NLR signaling pathway. This could shed some light on the potential therapeutic effect of SLC12A3, which is a SIMAP similar gene to SLC12A8 \cite{rattei2010simap}. Turning attention to the last disease in the table, myxedema, we discovered no direct literature evidence for the predicted genes, however, the latest UK Biobank database \cite{ukbiobank} prioritised IFNA2 as likely causal gene for myxedema through genome-wide associated study, which is missing in the evidence tensor we have built from Open Targets and PharmaProjects. Of course this initial superficial assessment of the predictions highlights the challenge of interpretability of such approaches, in which improvements will help us better understand the value and risks in such predictions. Further work is planned in this area.

\begin{table*}[]
    \centering
    \caption{The top 3 predicted gene associations for diseases with only a single gene target in the source data.}\label{tab:disease_low}
    \begin{tabular}{l c}
        \toprule
        \textbf{Disease}                                          & \textbf{Predicted Gene Targets} \\
        \midrule \midrule

        \multirow{3}{*}{Breast Adenosis (EFO:0006891)}            & SCN2A                           \\
                                                                  & HCN2                            \\
                                                                  & TYMP                            \\
        \midrule
        \multirow{3}{*}{Breast Fibrocystic Disease (EFO:0003014)} & SLC12A3                         \\
                                                                  & TYMP                            \\
                                                                  & FLT1                            \\
        \midrule
        \multirow{3}{*}{Myxedema (EFO:1001055)}                   & CHRM4                           \\
                                                                  & IFNA2                           \\
                                                                  & CSF3                            \\
        \bottomrule
    \end{tabular}

\end{table*}

\textbf{High Data Diseases.} To gain further insights from the predictions, we also investigated diseases with a larger number of connections in the source data and present novel associations predicted by the model in Table \ref{tab:disease_high}. Again some of the predicted associations do seem to be plausible. For example, two of the frequently occurring diseases in our dataset are cancers (prostate and lung), both for which the model ranks highly the gene CYP3A4. CYP3A4 is involved in the metabolism of, among other things, steroid hormones including testosterone which has lead to being identified as a potential factor in prostate cancer \cite{zeigler2001cyp3a4, chen2012roles}. HSD3B2 is also suspected to be involved in prostate cancer due to its similar role in steroid biosynthesis \cite{neubauer2018up}. CYP3A4s involvement in the metabolism of carcinogens has lead to suggestions of involvement with lung cancer \cite{jia2020cyp3a4, dally2003cyp3a4}. FGFR4 is a tyrosine kinase receptor and has been linked to poor prognosis in lung cancer by activating key pro-tumorigenic genes~\cite{quintanal2018fgfr4, quintanal2019fgfr4}. Moreover, the gene SGLT2, more commonly associated with metabolic diseases like diabetes, has also been hypothesised as being a way to reduce prostate cancer tumour growth by affecting glucose levels \cite{wright2020sglt2}. Turning attention to Type 2 Diabetes, it is interesting to notice that the PARP (poly[ADP-ribose] polymerase) enzyme family has a close link with diabetic cardiovascular dysfunction, e.g., its inhibition suppresses the development of diabetic endothelial dysfunction, and restores normal vascular function in established diabetes~\cite{Szabo2002parp} and PARP1 and PARP2 activity might have causal roles in the development of obstructive coronary artery disease \cite{Cui2020poly}. A recent animal study showed that Nischarin expression is attenuated in adipose tissue with obesity and that Nischarin mRNA has negative correlation with processes including obesity, fat distribution, lipid and glucose metabolism, suggesting that the NISCH gene could be a potential novel target for obesity-associated diabetes~\cite{Dong2019develop}.

\begin{table*}
    \centering
    \caption{The top 3 predicted gene associations for diseases with a large number of targets in the source data.}\label{tab:disease_high}
    \begin{tabular}{l c}
        \toprule
        \textbf{Disease}                                         & \textbf{Predicted Gene Targets} \\
        \midrule \midrule

        \multirow{3}{*}{Lung Cancer (EFO:0001071)}               & FGFR4                           \\
                                                                 & CUBN                            \\
                                                                 & CYP3A4                          \\
        \midrule
        \multirow{3}{*}{Type II Diabetes Mellitus (EFO:0001360)} & PARP2                           \\
                                                                 & PARP3                           \\
                                                                 & NISCH                           \\
        \midrule
        \multirow{3}{*}{Prostate Cancer (EFO:0001663)}           & CYP3A4                          \\
                                                                 & HSD3B2                          \\
                                                                 & SGLT2                           \\
        \bottomrule
    \end{tabular}
\end{table*}
\section{Discussion and Conclusion}\label{sec:conclusion}

To help facilitate the identification of potential drug targets at the early stage of drug discovery and development, in this paper we processed gene target-disease evidence data from the Open Targets platform as well as  clinical trial data from PharmaProjects database and employed standard machine learning classification algorithms and both matrix and tensor factorisation techniques to predict the clinical outcomes of unseen gene target-disease pairs. We adopted a new scheme when training the tensor factorisation method, i.e., alongside with a dense neural network, and showed that this outperforms many baseline models. Another main finding is that by using gene target representations learned from biomedical knowledge graphs  as the side information, the tensor factorisation performance can be dramatically improved, compared with traditional approaches where the side information only contains limited gene expression or protein-protein interaction data. 

There are a number of limitations of the work reported. First, we only used gene target embeddings learned from the knowledge graph as side information, the embeddings for diseases are not incorporated in the tensor factorisation framework. This is mainly due to the  fact that the drug discovery-oriented knowledge graph we have used in this analysis, Hetionet, uses a different disease ontology than the ones in Open Targets and PharmaProjects, and therefore has a rather small number of diseases (137 in total) -  as  a result, many diseases in our data tensor do not have corresponding embeddings. An obvious approach to address this would be to select or build a larger knowledge graph that contains all the diseases in the tensor and then generate the disease embeddings and use as the disease side information in the proposed framework. Second, we did not distinguish between various types of negative samples (i.e., failed gene target-disease clinical trials) in the collected data. In reality, the data negativeness is highly variable – multiple reasons could be responsible for the failure of a clinical trial: drug safety (toxicology or clinical safety), drug efficacy (failure to achieve sufficient efficacy), pharmacokinetics/pharmacodynamics (PK/PD), commercial benefits, company strategy, etc \cite{morgan2018impact}. These complex scientific and non-scientific factors make it difficult to obtain the true negative samples. In the future it would therefore be helpful to take this into consideration and classify the negative samples in the training set. Moreover, we treated all the positive samples, i.e., successful clinical trials, equally important in the training, however in practice, there exists some data that is more trustworthy than others. Hence, it would be important to take this into account when training the model by assigning different weights to different positive data points and penalise more if the model predicts the highly weighted data incorrectly.

Our work could be extended in several ways. For instance, it would be interesting to take into consideration more biological data sets such as ChEMBL to further enrich the gene target-disease-evidence tensor, which will help reduce the data sparsity. Another interesting line of investigation would be to design a more rigorous benchmarking strategy such as using well-curated disease databases for a handful of diseases of interest. Finally, the prediction improvements shown by utilising graph embeddings suggests there is still much to explore at the intersection between data modelling and graph composition, and algorithm application. 

\section*{Acknowledgement}

The authors would like to thank Ufuk Kirik, Manasa Ramakrishna, Natalja Kurbatova, Elizaveta Semenova and Claus Bendtsen for help and feedback throughout the preparation of this manuscript. We would also like to thank the OpenTargets team for their excellent resource. Stephen Bonner is a fellow of the AstraZeneca post-doctoral program.

\bibliographystyle{plain}
\bibliography{refs}

\end{document}